\documentstyle[12pt]{article}
\begin{document}
\begin{titlepage}
\null
\begin{center}
\hfill CERN-TH.7531/94\\
\hfill hep-ph/9412297
\end{center}
\vskip 1cm
\vfil
  \vskip .1in

\center{\Large\bf Supersymmetric Unification}
{}~\footnote{Talk
presented at the International Conference on the History of Original
Ideas and Basic Discoveries in Particle Physics held at Ettore
Majorana Centre for Scientific Culture, Erice, Sicily, July 29-Aug.4
1994.}

 \vskip .5in

\center{{\sc Savas Dimopoulos}~\footnote{On leave of absence
from the Physics Department, Stanford University, Stanford Ca. 94305,
USA. Work supported in part by NSF Grant NSF-PHY-9219345}\\{\it
Theory Group, CERN}\\{\it 1211 Geneva 23, Switzerland}}

\vskip .5in

\begin{abstract}
The measured value of the weak mixing angle is, at present, the only
precise experimental indication for physics beyond the Standard
Model. It points in the direction of Unified Theories with
Supersymmetric particles at accessible energies. We recall the ideas
that led to the construction of these theories in 1981.
\end{abstract}
\end{titlepage}

\section{Why Supersymmetric Unification?}

It is a pleasure to recall the ideas that led to the first
Supersymmetric Unified Theory and its low energy manifestation, the
Supersymmetric $SU(3)\times SU(2)\times U(1)$ model. This theory
synthesizes two marvelous ideas, Unification \cite{GUTs} and
Supersymmetry \cite{SUSY,sugra}. The synthesis is catalyzed by the
hierarchy problem \cite{hier} which suggests that Supersymmetry
occurs at accessible energies \cite{hierSUSY}. Since time is short
and we are explicitly asked to talk about {\it our own} contributions
I will not cover these important topics.

A look at the the program of this Conference reveals that most other
topics covered are textbook subjects, such as
Renormalization of the Standard Model \cite{thooft} and Asymptotic
Freedom \cite{gross}, that are at the foundation of our field. So it
is natural to ask why Supersymmetric Unification is included in such
a distinguished company of well-established subjects? I am not
certain. Perhaps the main reason at present is a quantitative {\bf
prediction}, dating from 1981, that has been verified by high
precision data
(see figure); that is a correlation between $\alpha_s(M_Z)$ and
$\sin^2(\theta_W)$ which has been confirmed by experiment at the 1\%
level \cite{lp}. In fact this is the only significant {\it
quantitative} success of any extension of the Standard Model, and is
the
strongest experimental hint that we have for physics beyond the
Standard Model .

\begin{figure}[t]
\vspace{11.16cm}
{\bf Figure:}\ \ {\small The correlation in the values of
$\sin^2(\theta_W)$ and $\alpha_s(M_Z)$ predicted in SUSY-GUTs and
ordinary GUTs. The bare Superstring prediction is the point on the
far right. The present 1994 data are contrasted with the 1981 data.
The bands are the uncertainties in the theoretical predictions of
GUTs and SUSY-GUTs. The numbers in the bands indicate the Unification
Scale. The uncertainties in the theoretical predictions for
superstrings are not known.}
\end{figure}

It is amusing that this prediction of the weak mixing angle, at the
time it was made, disagreed with experiment \cite{swogu} (see
figure). Experiments since then, especially the recent LEP results,
have convincingly changed the experimental world average
 in favor of the prediction. The success of this
prediction depends crucially on having both Unification {\bf and} low
energy Supersymmetry in the same theory; either Unification or
Supersymmetry alone are insufficient. So, although we have not seen
any superparticles yet, we have evidence for Superunification !

We will discuss the developments in chronological order, beginning
with the state of our field before 1981.

\section{Before  1981.}

A crucial turning point in our field occurred in the Spring of 1978.
The SLAC experiment on parity violation in neutral currents convinced
many theorists that the Standard Model of Glashow, Weinberg and Salam
was correct and that it was a good time to start focusing on the next
layer of questions: to explain some of the puzzling features of the
Standard Model. The first question that theorists turned to was the
``hierarchy problem'' \cite{hier}: attempting to understand why the
Higgs
mass is so much smaller than the Planck mass or the Unification
Scale. The Higgs does not carry any symmetry that ensures its
lightness; indeed, in the absence of miraculous cancellations, the
Higgs mass would be driven to the Planck or unification scale; it
would not be available at low energies to do its intended job of
giving mass to the weak gauge bosons.

 Susskind and Weinberg \cite{ws} proposed the very appealing idea of
Technicolor, as an alternative to the Higgs, for giving mass to the
weak gauge bosons. In early '79 Technicolor was enlarged into
``Extended Technicolor'' \cite{etc} to allow the quarks and leptons
to get their masses. By the summer of 1980 it became clear that these
theories suffered from generic problems of flavor violations
\cite{je} that could perhaps be cured only by complicating the theory
immensely and losing any hope of calculability. I, perhaps
prematurely, felt that this was too high a price to pay and decided
to look at other alternative approaches to the Hierarchy problem.

That is when we turned to Supersymmetry \cite{SUSY,sugra}. It was
generally realized that Supersymmetry could help the hierarchy
problem \cite{hierSUSY}. The reason is that the Higgs, a scalar,
would form a degenerate pair with a fermion, called the Higgsino.
Since the Higgsino could be protected by a chiral symmetry from
becoming superheavy, so could its degenerate scalar partner, the
Higgs. Of course Supersymmetry does much more than to just relate the
Higgs to the Higgsino. It assigns a degenerate scalar
``superpartner'' to each and every known quark and lepton, as well as
a fermionic degenerate superpartner to each gauge boson. Since no
such particles had been seen it was clear that Supersymmetry had to
be a broken symmetry. Nevertheless, Supersymmetry would still help
the hierarchy problem as long as its breaking occurs near the weak
scale, the superpartners
are at accessible energies !
This line of reasoning led us to begin our attempt to find a
Supersymmetric version of the Standard Model with Supersymmetry
broken at the weak scale. Together with Stuart Raby and  Leonard
Susskind we started learning about Supersymmetry and tried to find
out if such theories had already been constructed. We quickly
discovered that no Supersymmetric versions of the Standard Model
existed at that time. There were early attempts by Fayet \cite{fayet}
that were plagued by the following problems:
\begin{itemize}
\item They were {\bf not} Supersymmetric extensions of the Standard
$SU(3)\times SU(2)\times U(1)$ model. The gauge group was
$SU(3)\times SU(2)\times U(1)\times U(1)'$. Without the extra $U(1)'$
neutral current interactions, the theory had phenomenological
problems: it spontaneously violated electric charge and color
conservation.

\item The extra neutral current $U(1)'$ was anomalous. To get a
consistent theory one had to cancel the anomalies by introducing
extra light particles, distinct from the ordinary quarks and leptons.
These again led to color and electric charge breaking at the weak
scale: the photon had a mass $\sim M_W$.\footnote{There were other
problems such as a  continuous R-symmetry which forced the gluino to
be massless.}
\end{itemize}
The root of these problems was that in these theories Supersymmetry
was broken spontaneously at the tree level. In 1979 a very important
paper by Ferrara, Girardello and Palumbo \cite{fgp} showed that in
such theories, under very general conditions, color and charged
scalars would get negative masses squared, leading to breaking of
electric charge and color. This essentially stopped efforts to build
realistic Supersymmetric theories.

During the fall of 1980 these difficulties often caused us to wonder
whether we were not on the wrong track and Supersymmetry was doomed
to fail in the real world, at least as a low energy phenomenon.
Almost nothing that we tried worked. We could not cancel the
anomalies of the extra $U(1)'$ without giving a mass $\sim M_W$ to
the photon and gluon. The extra $U(1)'$ made it impossible to
unify\footnote{This was not a problem for the early
attempts \cite{fayet}; Unification was not one of their objectives
since they had not associated Supersymmetry with the hierarchy
problem.}, which is important for addressing the hierarchy problem
and for predicting $\sin^2(\theta_W)$.
We also had no clue as to what to do with the continuous R-symmetry
that
implied massless gluinos. Little of what we learned during these
early exercises has survived. The benefit was that we learned bits of
the mathematical formalism which told us, loud and clear, that all
known particles have superpartners and how these couple\footnote
{Hypercharge anomalies dictated an even number of Higgs doublets.
Later, $\sin^2(\theta_W)$ forced the number of doublets to be 2
\cite{dg}. } .
However, given all the problems, it was unclear that anything that we
learned would survive.

Since the origin of these problems was  that   Supersymmetry was
broken  spontaneously,  it seemed clear
to us that we should look for alternate mechanisms to break
Supersymmetry.
At first, we attempted
 to break Supersymmetry dynamically
with a new strong force, very similar to Technicolor, which we called
Supercolor. We were not alone in these efforts. Witten
\cite{hierSUSY} as well as Dine, Fischler and Srednicki
\cite{hierSUSY} were pursuing similar ideas for precisely the same
reasons. They wrote two very important papers entitled ``Dynamical
breaking of Supersymmetry ''(Witten) and ``Supersymmetric
Technicolor'' (Dine, Fischler and Srednicki).   Their preprints
appeared in April of '81 at the same time as our ``Supercolor''
paper \cite{hierSUSY}.
 An essential objective of these works was to point out that
low energy Supersymmetry helps the hierarchy problem\footnote{We
were, for sure, not alone in being aware of this. Several theorists,
in addition to those in Reference \cite{hierSUSY}, knew it and did
not
publish it. The problem was to implement the idea in a  consistent
theory, incorporating Supersymmetry at low energies.}, and to argue
that a new strong force analogous to QCD or Technicolor may induce
the breaking of Supersymmetry and explain the smallness of the
electroweak scale. Dine, Fischler and Srednicki, as well as Raby and
myself, also attempted to build explicit models incorporating these
ideas, but without much success. I do not have time to discuss
these ``Supercolor'' or ``Supersymmetric Technicolor'' theories. They
had  problems; one of them was that they were baroque. By January of
1981 we were very discouraged. Although Stuart Raby and I had begun
writing the Supercolor paper \cite{hierSUSY}, we already did not
believe in it. It seemed too much to believe that Nature would make
simultaneous use of Supersymmetry {\bf and} Technicolor to solve the
hierarchy problem.

\section{1981.}
\subsection{Distancing Ordinary Particles from the Origin of
Supersymmetry Breaking.}

In January of 1981 the prospects for a realistic
Supersymmetric model were not bright. Models with spontaneusly broken
Supersymmetry had grave phenomenological problems. Dynamical
Supersymmetry breaking models were at best baroque. {\it The
prevailing view was that a realistic Supersymmetric model would not
be found until the problem of Supersymmetry Breaking was solved.  It
was further believed  that the experimental consequences of
Supersymmetric theories would strongly depend on the details of the
mechanism of Supersymmetry breaking.} After all, it was this
mechanism that caused the phenomenological disasters of the early
attempts.

The key that took us out of this dead end grew out of our protracted
frustration with the above problems, and our desire to do physics
with
the idea of superunification. These --- quite suddenly --- led us to
switch problems, adopt  a more phenomenological approach and simply
assume that the dynamics that breaks Supersymmetry is external to and
commutes with the ordinary $SU(3)\times SU(2)\times U(1)$ sector;
specifically, we postulated that:
\begin{enumerate}
\item The only particles carrying $SU(3)\times SU(2)\times U(1)$
quantum numbers are the ordinary ones and their Superpartners that
reside at the weak scale. Extra particles with exotic  $SU(3)\times
SU(2)\times U(1)$ quantum numbers are unnecessary.
\item The ordinary particles and their superpartners do not carry any
extra new gauge interactions at low energies. This is essential for
SU(5) unification.
\item The sole effect of the Supersymmetry breaking mechanism is to
lift the masses of the supersymmetric partners of all ordinary
particles to the weak scale.
\end{enumerate}
These hypotheses helped us sidestep the obstacles that stood in the
way of Unifying and doing physics in the ordinary $SU(3)\times
SU(2)\times U(1)$ sector, the domain of experimental physics ! The
question of the origin of Supersymmetry Breaking had been
circumvented; a good thing, since this question continues to remain
open.
These hypotheses started bearing fruits immediately.
\newpage
\subsection{Raby and Wilczek: \,``Supersymmetry and the Scale of
Unification.''}
In this paper \cite{drw1} we computed the Unification Mass when you
have a minimal Supersymmetric particle content. We found that,
because the superpartners of the gauge bosons slow down the evolution
of the couplings, the unification mass increased to about $10^{18}$
GeV. This was interesting for two reasons:
\begin{itemize}
\item This value is close to the Planck mass, perhaps suggesting
eventual unity with gravity\footnote{This connection got weaker as
more accurate calculations \cite{lp,dg,sin2} reduced the value to
$\sim 2\times 10^{16}\ GeV$. }.
\item There was a distinct experimental difference with ordinary
$SU(5)$: the  proton lifetime was unobservably long.
\end{itemize}
The latter appeared to
be an easily disprovable prediction. In fact by that time
three different experimental groups had reported preliminary proton
decay ``candidate events'': the Kolar gold field, Homestake mine and
the Witwatersrand experiments. We knew that S.Miyake, of the Kolar
Gold Field experiment, and possibly representatives of the other
experiments were going to talk about their events in the upcoming
``Second Workshop on Grand Unification'' where I was also going to
present our theoretical results. So, I was a bit nervous but did not
hesitate for a moment to present them. I was, and still am, very
proud of this paper. A simple and well motivated ingredient, virtual
superparticles, made a huge difference to a quantity that was being
measured at that time, the proton lifetime. Perhaps this is the first
test that SUSY-GUTs have passed.
In this paper, although we pointed out that the value of
$\sin^2(\theta_W)$ would change due to the Higgs sparticles, we did
not present the new value. After satisfying ourselves that it would
not be grossly modified, we focused
on the change in the unification mass, which at that time was more
important for experiment.

The next big step was to construct a realistic supersymmetric theory.
\newpage
\subsection{Georgi: \,`` Supersymmetric GUTs.''}
These two papers  \cite{dg} titled  ``Softly Broken Supersymmetry and
SU(5)''   and `` Supersymmetric GUTs'' accomplished three objectives:

\begin{enumerate}
\item {\bf Supersymmetric Unification (SUSY-GUTs):}
 Construction of a Unified supersymmetric theory of strong and
electroweak forces. Our gauge group was SU(5). This was not really
much harder than building a non-unified theory. Unification was also
essential for the prediction of $\sin^2(\theta_W)$ and for some of
the phenomenology, such as  proton decay and gaugino masses. It was
also a good framework for addressing the hierarchy problem.
\item {\bf Supersymmetry Breaking:}
Supersymmetry was broken softly but explicitly by mass terms for all
scalar superpartners and gauginos. The origin of supersymmetry
breaking was not specified. As long as it is external to and commutes
with $SU(3)\times SU(2)\times U(1)$ it does not matter for
experiment\footnote{See Sections 3.1 and 5.}. Ingenious ideas for
generating these soft terms by either new gauge forces \cite{hs}, or
via supergravity \cite{suGUT} were proposed a year later.
\item {\bf Supersymmetric Standard Model (SSM):}
As a bonus, our theory contained the first phenomenologically viable
supersymmetric extension of the standard $SU(3)\times SU(2)\times
U(1)$ model (SSM).

\end{enumerate}
We constructed the model in late March and early April of 1981. We
were overjoyed. We had the first realistic Supersymmetric theory,
incorporating all non-gravitational phenomena and valid up to the
Planck mass. We immediately started thinking about experimental
consequences. We wanted to make sure that  we would not miss anything
important. Time pressure helped us a lot. Both Howard and I were
scheduled to give two consecutive talks in the Second Workshop on
Grand Unification which took place at the University of Michigan on
April 24-26, 1981. Here are some of our phenomenological results that
we reported in that Workshop \cite{dg}:
\newpage
\begin{itemize}
\item {\bf  $\sin^2(\theta_W)$ :}
We presented our SUSY-GUT prediction for $\sin^2(\theta_W)$. The
magnitude we got disagreed with the central experimental value, but
the errors were large. We argued that there would have to be 2 Higgs
doublets for the value not to be too far off.
\item {\bf Proton Decay}: We reported that the Supersymmetric
Unification Mass is so large \cite{drw1} that proton decay is
unobservably small.
\item {\bf Superparticle Spectroscopy: squarks and sleptons. }
We postulated that all squarks and sleptons have a common universal
mass ($\sim M_W$) at the unification scale. This way we had a
Super-GIM mechanism supressing flavor violations. The Higgses had
different masses.
\item {\bf Superparticle Spectroscopy: gauginos. }
Because we had a unified theory all gauginos had a common Majorana
mass ($\sim M_W$) at the unification scale.
\item {\bf Family Reflection Symmetry; Stable LSP. }
To avoid rapid proton decay via dimension-four operators we
postulated a discrete symmetry forbidding three-family couplings.
This symmetry was subsequently called family reflection
symmetry\cite{drw2} or matter parity. We concluded:
\\[2mm]
 {\it ``the
lightest of the supersymmetric particles is stable. The others decay
into it plus ordinary particles. One simple possibility is that it is
the U(1) gauge fermion.''}\footnote{Continuous symmetries, such as
Technibaryon number \cite{ws} or continuous R-symmetry \cite{fayet}
also lead to new stable particles. In fact, many extensions of the SM
have this feature.}
\end{itemize}

It is gratifying that the above ingredients have survived the test of
time. They form the basis of what is now called the minimal
supersymmetric standard model (MSSM) \footnote{The acronym MSSM is
often used incorrectly to mean minimal SUSY-GUTs. Gell-Mann's terms
``Superstandard model'' and ``Superunified theory'' are much better
but not widely used.}. Perhaps the most important conclusion of our
paper is also the one that now seems so evident because it has, with
time, been incorporated into our thinking:
\newpage
\begin{quote}
{\it ``The phenomenology of the model is simple. In addition to the
usual light matter fermions, gauge bosons and Higgs bosons, we
predict heavy matter bosons, gauge fermions and Higgs fermions as
supersymmetric partners. We can say little about their mass except
that they cannot be very large relative to 1 Tev or the motivation
for the model disappears.''} \cite{dg}
\end{quote}
Of course, our motivation was to address the hierarchy problem;
without it we could not have drawn this conclusion.

Georgi and I spoke on the last day of the conference \cite{swogu}. My
feeling was that our results were for the most part ignored,
especially by the experimentalists who did not care about the
hierarchy problem. Our conclusions were very much against the spirit
of the conference. There were three things against us:
\begin{itemize}
\item The central value of the weak mixing angle agreed better with
the predictions of ordinary (non-Supersymmetric) Grand Unified
Theories, albeit with large error bars (see figure).
\item Preliminary proton decay ``candidate events'' had been reported
by three different experimental groups, the Kolar gold field,
Homestake mine and the Witwatersrand experiments.
\item The host institution was gearing up to launch the biggest
effort on proton-decay namely the IMB experiment.
\end{itemize}
The atmosphere in the conference is summarized by Marciano's April
24, 1981 concluding remarks \cite{swogu}:
\begin{quotation}
``The basic idea of Grand Unification is very appealing. The simplest
model based on $SU(5)$ has scored an important success in predicting
a value for {\bf $\sin^2(\theta_W)$ which is in excellent agreement
with recent experimental findings} (after radiative corrections are
included). It makes an additional dramatic prediction that the proton
will decay with a lifetime in the range of $10^{30}$--$10^{32}$
years. If correct, such decays will be seen by the planned
experiments within the coming year ({\bf or may have already been
seen}). An incredible discovery may be awaiting us.''\footnote{The
emphasis here is mine.}
\end{quotation}

It is remarkable that Georgi, in such an atmosphere, did not hesitate
to propose an alternative to his and Glashow's '74 theory \cite{GUTs}
which seemed to be on the verge of being proven. But this is Howard !
He would rather have fun with physics than worry about such things.

Very significant encouragement came from Sheldon Glashow, Leonard
Sus\-skind and Steven Weinberg. In his April 26, 1981 conference
summary talk \cite{swogu} Weinberg  mentioned our theory and its
predictions of $\sin^2(\theta_W)$ and $M_{GUT}$ several times.   His
verdict \cite{swogu}:
\begin{quote}
{\it ``...the model of Dimopoulos and Georgi has many other
attractive features and something like it may turn out to be
right.''}
\end{quote}

This was music to my ears.\\

In May I presented our results in two more conferences, one in Santa
Barbara and the other at the Royal Society in London. Soon afterwards
theoretical activity in supersymmetric unification began to pick up.
In August of '81 Girardello and Grisaru wrote a very important paper
\cite{gg} systematically discussing explicit soft breaking of global
supersymmetry; they were the first to discuss cubic soft terms.
Starting in July of '81 several important papers \cite{sin2} with
 the calculation of the superunified value of  $M_{GUT}$ and
$\sin^2(\theta_W)$ appeared, some improving it to two loops.
 Sakai's paper \cite{sin2} includes an analysis of SU(5) breaking
which is very similar to ours; it does not introduce the soft
superparticle
mass terms that break supersymmetry and thus does not address the
phenomenology of superparticles.

The interest in GUTs and SUSY-GUTs dwindled after 1983. The rise of
superstrings, the absence of proton decay and the lack of precise
data on $\sin^2(\theta_W)$ were some of the reasons. The morale among
the non-stringers was so low that the annual series of ``Workshops on
Grand
Unification'' was terminated.  1989 was the year of the ``Last
Workshop on Grand Unification''. In the introduction to that terminal
volume Paul Frampton exclaimed:\\[2mm] {\it `` Alas, none of the
principal
predictions of GUTs have been confirmed.''}\\[2mm] This was written
in
August 1989, just as LEP was beginning to take data...

\section{Proton Decay Revisited.}
Although Georgi and I worried a lot about dimension-four baryon
violating operators and we introduced the family reflection symmetry
to forbid them, it did not occur to us to check the operators of
dimension five ! Weinberg \cite{wsy} as well as Sakai and Yanagida
\cite{wsy} studied these operators and concluded that they pose a
severe problem for our theory. They attempted to construct models
with an extra $U(1)'$ gauge group that would forbid the dimension
five operators that mediated proton decay. Raby, Wilczek and I
studied these operators in October of '81 and concluded  that the
small Yukawa couplings of the light generation naturally supressed
these operators \cite{drw2}. The resulting proton decay rates,
although not calculable from low energy physics parameters, could be
experimentally observable. Furthermore they had a very unique
signature that is not expected in non-supersymmetric theories:
protons and neutrons decay into kaons.  We were very excited that we
had identified another ``smoking gun'' for supersymmetry. Ellis,
Nanopoulos and Rudaz independently reached the same conclusions
\cite{drw2}.
\section{Completing the Picture.}
Since time is so short I have limited myself to those aspects of
superunified theories that are least model-dependent and
experimentally testable or, in the case of $\sin^2(\theta_W)$ and
proton decay, perhaps already tested. Of course, the theory that we
proposed is far from complete and left many important theoretical
questions unanswered. I will briefly mention some of the problems and
related ideas.

{\bf Doublet-triplet splitting:} There is one remaining technically
natural fine tuning in our theory \cite{dg}. Wilczek and I addressed
this problem in June of 1981 and found two solutions now called the
missing partner and the missing VEV mechanisms \cite{dw}. Attempts to
implement these mechanisms in realistic theories led to very
complicated constructions \cite{gr}. This continues to be an open
problem.

{\bf Hidden sector:} The theoretical question of how supersymmetry is
broken and superparticle masses are generated in our theory attracted
a lot of attention.  Georgi and I had spent a couple of days thinking
about this and then decided that it was not phenomenologically
interesting: {\it two different theories of supersymmetry breaking
that give precisely the same soft masses to the superpartners of
ordinary particles cannot be experimentally distinguished}. So we
abandoned it. Our philosophy was to build an effective theory that
describes the $SU(5)$ part of supersymmetric world, which {\it is}
accessible to experiment. It could result from many different ways of
breaking Supersymmetry, as long as the postulates of Section 3.1 are
satisfied.

Nevertheless, it was important to present at least an existence proof
of a mechanism that generated our soft terms. An important
consideration was that squarks and sleptons belonging to different
generations had to have identical masses to avoid problems with rare
processes \cite{dg}. In the winter/spring of '82 three different
groups \cite{hs}, Dine and Fischler, Raby and I, and Polchinski and
Susskind came up with the idea of a Hidden Sector, around $10^{11}$
GeV,  where supersymmetry breaking originates and is subsequently
communicated to the ordinary particles via a new gauge interaction at
the unification scale\footnote{For Raby and me the starting point was
trying to build a realistic model utilizing Witten's idea of
``Inverted Hierarchy''  \cite{w}. }. Soon afterwards a series of very
important papers developed a better idea for such a mechanism:
Supersymmetry breaking could be communicated from the hidden sector
via supergravity \cite{suGUT}.

{\bf Radiative electroweak breaking:}  Hidden sector mechanisms for
Supersymmetry breaking, under very special assumptions, give
degenerate masses to all scalars: squarks, sleptons as well as
Higgses. This is good for avoiding flavor violations \cite{dg} but
poses the puzzle: what distinguishes the Higgs from the squarks and
the sleptons? Why does the Higgs get a vacuum expectation value and
not the squarks?\footnote{In the original SUSY-GUT this was not an
issue because the Higgs masses were assumed to be different from the
universal squark and slepton masses \cite{dg}.}. Starting with
Iba\~nez and Ross, a series of very important papers  \cite{rewsb}
developed the idea of radiative electroweak breaking which answers
this question dynamically provided the top quark is sufficiently
heavy, above $\sim 60$  GeV.

The title of this section is misleading. The picture is still very
far from complete; many fundamental questions remain unanswered. The
theory we have is definitely {\bf not} a theory of everything.
Instead, it is a phenomenological, disprovable theory that allows us
to make contact with experiment in spite of the questions that it
fails to address.
\section{Prospects.}
In the last few minutes I want to take a break from history and
mention some contemporary issues.

 \subsection{ How significant is the $\sin^2(\theta_W)$ prediction? }
\begin{table}[t]
\vspace{4.82cm}
{\bf Table:}\ \ {\small The experimental values for
$\sin^2(\theta_W)$ and $\alpha_s(M_Z)$ are contrasted with the
predictions of three theories: Ordinary GUTs, SUSY-GUTs and bare
Superstrings . Under each prediction we list the number of standard
deviations that it differs from experiment. GUTs and SUSY-GUTs
predict one of either $\sin^2(\theta_W)$ or $\alpha_s(M_Z)$; the
other one is an input.  For Strings both $\sin^2(\theta_W)$ {\it and}
$\alpha_s(M_Z)$ are predictions. The uncertainties in the theoretical
predictions for superstrings are not known.}
\end{table}

 {\bf SUSY-GUTs:} Since the LEP data confirmed the SUSY-GUT
prediction this topic has received a lot of attention and is
discussed in many papers. Excellent recent analyses are those of
Ref. \cite{lp}. We summarize the results in the table and the figure
which together with their captions tell the story. The estimated
uncertainties in the theoretical predictions for SUSY-GUTs and GUTs
are due to: $\alpha_s(M_Z)$ and $\alpha(M_Z)$ error bars, sparticle
thresholds, $m_t$ and $m_{h^0}$, GUT thresholds and
Non-renormalizable operators at the unification scale.  For the
$\sin^2(\theta_W)$ prediction they all add up to about $\pm 1\%$
\cite{lp}\footnote{$\sin^2(\theta_W)$ is in the $\overline {\rm MS}$
scheme. }.
 The experimental error is negligible, $\pm 0.2\%$. The
probability that the agreement is an accident is $\sim 2\%$. The
largest source of theoretical uncertainty is due to the
$\alpha_s(M_Z)$ error bar; this should shrink in the future. The
other uncertainties are significantly smaller. The threshold
corrections are proportional to $\alpha$s times logarithms of mass
ratios. For example, the total of the low energy sparticles'
contributions is summarized in the following elegant expression
\cite{lp,cpw}:
\begin{equation}
\sin^2\theta(M_Z) = 0.2027 + \frac{0.00365}{\alpha_3(M_Z)}
- \frac{19 \alpha_{em}(M_Z)}{60 \pi} \ln\left(\frac{T_{SUSY}}
{M_Z}\right)
\end{equation}
where\footnote{In eq.(2) if any mass is less than $M_Z$ it should be
replaced by $M_Z$. },
\begin{equation}
T_{SUSY} =
m_{\widetilde{H}}
\left( \frac{m_{\widetilde{W}}
}{m_{\tilde{g}}}
\right)^{28/19}
\left[
\left( \frac{m_{\tilde{l}}}{m_{\tilde{q}}}
\right)^{3/19}
\left( \frac{m_H}{m_{\widetilde{H}}}
\right)^{3/19}
\left( \frac{m_{\widetilde{W}}}{m_{\widetilde{H}}}
\right)^{4/19} \right] .
\label{eq:SUSYm}
\end{equation}
and $m_{\tilde{q}}$,
$m_{\tilde{g}}$,
$m_{\tilde{l}}$,
$m_{\widetilde{W}}$,
$m_{\widetilde{H}}$ and $m_H$
are the characteristic masses of the
squarks, gluinos, sleptons, electroweak gauginos,
Higgsinos and  the heavy Higgs doublet,
respectively. $T_{SUSY}$ is an effective SUSY threshold.

{}From these equations we learn that the supersymmetric threshold
corrections are typically small. The same holds for the high energy
threshold corrections in minimal SUSY--GUTs \cite{lp}. Therefore the
$\sin^2\theta(M_Z)$ prediction is quite  insensitive to the details
of both the low and the high mass-scale physics; it takes a number of
highly split multiplets to change it appreciably. For example, we
know that to bring $\sin^2\theta(M_Z)$ down by just $\sim 10\%$  ---
back to the standard SU(5) value --- we would need to lift the
higgsinos and the second higgs to $\sim 10^{14} GeV$.

The flip side of these arguments show that to ``fix'' Standard GUTs,
you also need  several highly split multiplets \cite{fg}. In fact you
need many more, since you do not have superpartners.
The figure and the table show that in Standard GUTs either
$\sin^2(\theta_W)$ or $\alpha_s(M_Z)$  are off by many standard
deviations. Worse yet, the proton decays too fast. Do these problems
mean that all non-supersymmetric GUTs are excluded? Of course not. By
adding many unobserved split particles at random to change the
running of the couplings you can {\it accommodate} just about {\bf
any} values of $\sin^2(\theta_W)$ and $M_{GUT}$. So, in what sense
are these quantities {\it predicted}\,?\\

I answer this with a quote from Raby and Wilczek \cite{phystoday}:
\begin{quote}
{\it`` Once we wander from the straight and narrow path of
minimalism,
infinitely many silly ways to go wrong lie open before us.
In the absence of
some additional idea, just adding unobserved particles at random to
change
the running of the couplings is almost sure to follow one of these.
However there are a {\bf few ideas} which do motivate definite
extensions of
the minimal model, and {\bf are sufficiently interesting that
even their failure
would be worth knowing about}.''}\footnote{Emphasis mine}
\end{quote}
The '81 predictions for $\sin^2(\theta_W)$ and $M_{GUT}$ were
inevitable consequences of an {\it idea}; they could not be modified,
although they came at a time when they were least expected and, for
sure, unwanted.

{\bf  Peaceful coexistence with Superstrings}: The predictions that
we quote in the table for superstrings assume the minimal
supersymmetric particle content up to the string scale $M_s$ of about
$4\times 10^{17}$ GeV, and do not include any potentially large
string-induced corrections\footnote{The bare string value of
$\alpha_s(M_Z)=0.2$ gives a proton mass of approximately 20 GeV. This
follows
directly from $M_s/M_{GUT} \simeq 20$.}. These corrections are model
dependent; in the absense of a model, it is not possible to estimate
their magnitude. It is clear that the corrections would have to be
quite large to make up for the large discrepancies with experiment .
It is possible that a model will be found  where the corrections are
large and can be tuned to accommodate the data. Such a ``fix'' seems
no better than accomodating ordinary SU(5) with large corrections
caused by random unobserved multiplets. Perhaps a more appealing
solution will be found. Such a solution should answer the question
posed by Barbieri et al. \cite{barb}:\\[2mm]
\newpage
{\it ``why should these corrections maintain the relations between
the couplings characteristic of the  Grand Unified symmetry, if such
a symmetry is not actually realised?''}\\[2mm]
One possibility is that at $M_s$ the string theory breaks to a
SUSY-GUT \cite{barb,strGUT} ; this is a  promising new direction
which may combine some of the virtues of both SUSY-GUTs and strings.
A challenge to such attempts would be to explain the ratio of the
SUSY-GUT scale to the string scale.

\subsection{Life before LHC}
{\bf Where are the Sparticles?}
The short answer is still:\\[2mm]
{\it ``...We can say little about their mass except that they cannot
be very large relative to 1 Tev or the motivation for the model
disappears.''} \cite{dg}\\[2mm]
Upper limits can be obtained which are functions of the amount
of fine tuning that you allow in the theory \cite{bg}, but this is
clearly a matter of taste. For any fixed amount of fine tuning  a
large top Yukawa coupling pushes down many sparticle masses; so these
upper
limits are now known to be near their minima. For example, if you
only allow 10\% fine tunings then the two lightest neutralinos and
chargino would be accessible at LEP 200.

Obviously, sparticle masses are proportional to the SUSY breaking
scale. In contrast, the lightest Higgs mass has {\it logarithmic}
sensitivity to the SUSY scale since its mass is proportional to the
weak VEV --- which of course is fixed, after the requisite fine
tuning. As a result, even if the sparticles are at $\sim 10$ TeV the
{\it upper} limit to its mass is only $\sim 160$ GeV
\cite{Higgs,cpw} ; it drops to $\sim 120$ GeV if the sparticles are
below a TeV.\footnote{In contrast to SUSY-GUTs, in the SM vacuum
stability gives a {\it lower} bound to the Higgs mass of $\approx
135$
GeV.}

If we are really lucky then,  before LHC is built, we may see
sparticles at LEP 200 or proton decay into kaons at SuperKamiokande
or Icarus. What if we are not? There are still some other possible
consequences of SUSY-GUTs. These have to do with rare processes.

{\bf Hall-Kostelecky-Raby effects:}
The postulated universality of the masses of sparticles belonging to
different generations \cite{dg} suppresses rare processes.
Universality means that the sparticle mass matrix is ``isotropic'' in
flavor space; only the quark masses spoil this isotropy by virtue of
their differing eigenvalues and $V_{KM}$. As a result, just like in
the SM, all flavor violating quantities involve the usual left handed
angles and phase of $V_{KM}$  and are under control. In GUTS there
are more physical angles and phases by virtue of transitions caused
by the extra gauge particles. In ordinary GUTs these do not matter at
low energies; they decouple like $M^{-1}_{GUT}$. Not so in
SUSY-GUTs\, !
\cite{hkr}. The sparticle
 masses are distorted by these extra angles and
phases and propagate new flavor violations down to low energies,
especially if there are large Yukawa couplings such as the top
quark's \cite{bh}.
These can lead to interesting flavour or CP violating
effects, even in minimal SUSY-GUTs,  such as $\mu \rightarrow e $
transitions \cite{bh} and electric dipole moments for the neutron
$(d_n)$ and electron $(d_e)$  \cite{dh}\footnote{These effects,
just like $\sin^2(\theta_W)$, are only logaritmically sensitive to
the GUT scale\, ! Unlike $\sin^2(\theta_W)$, they depend on powers of
the yet unknown sparticle masses.}. These already put interesting
bounds on sparticle parameters. Experimentalists are encouraged to
look for these effects in the near future.

These effects are expected to be even bigger, perhaps too big, in
theories that explain flavour \cite{g} suggesting that sparticles are
too heavy to be accessible at LHC. Is it possible to have light
sparticles and these effects suppressed? Yes. The soft sparticle
masses could be truly ``soft '' and disappear at high
energies\footnote{SUSY  breaking could be communicated by
particles much lighter than $M_{GUT}$ to the visible sector.};
 then they cannot sense the
extra flavor physics occuring at $M_{GUT}$. One would only give up
the usual lore where supersymmetry is fed from the hidden sector via
supergravity\footnote{In this case the soft masses are definitely
``hard'': they persist up to the Planck scale and they can sense the
extra flavor physics.}. This would change none of the experimental
 consequences of the '81 softly broken SUSY-GUTs.

\section{Acknowledgments}
I would like to thank: R. Barbieri, M.Carena, G.Giudice, N.Polonsky
and C. Wagner for very valuable conversations; Ann Georgi and her
family for having my family as houseguests during two very exciting
weeks in March and April of '81; Professor Harvey Newman and his
family, Professor Zichichi, Professor Ypsilantis and the friendly
people of Erice for organizing a marvelous conference and for their
hospitality.


\newpage
\begin{thebibliography}{99}
\bibitem{GUTs} H. Georgi and S. Glashow, Phys. Rev. Lett. 327 (1974)
438;\\ J. Pati and
A. Salam, Phys. Rev. D 8 (1973) 1240;\\ H. Georgi, H. Quinn and S.
Weinberg, Phys. Rev. Lett. 33 (1974) 451.

\bibitem{SUSY} Yu. A. Gol'fand and E.P. Likhtman, JETP
Lett. 13 (1971) 323;\\ D.V. Volkov and V.P. Akulov, Phys.
Lett. B 46 (1973) 109;\\ J. Wess and B. Zumino, Nucl.
Phys. B 70 (1974) 39.

\bibitem{sugra} D.Z. Freedman, P. van Nieuwenhuizen, and S. Ferrara,
Phys. Rev. B 13 (1976) 3214;\\ S. Deser and B. Zumino, Phys. Lett. B
62 (1976) 335.

\bibitem{hier} K. Wilson, as mentioned in L. Susskind, Phys. Rev. D
20 (1979)
2619;\\ E. Gildener, Phys. Rev. D 14 (1976) 1667;\\
E. Gildener and S. Weinberg, Phys. Rev. D 15 (1976) 3333.

\bibitem{hierSUSY}  L. Maiani, Proceedings of the Summer School of
Gif-Sur-Yvette (Paris 1980);\\ M. Veltman, Acta Phys. Polon.
B 12 (1981) 437;\\ S. Dimopoulos and S. Raby, Nucl. Phys. B 192
(1981)
353;\\  E. Witten, Nucl. Phys. B 188 (1981) 513;\\
M. Dine, W. Fischler, and M. Srednicki, Nucl. Phys. B 189 (1981)
575;\\ ibid., B 202 (1982) 238.

\bibitem{thooft} See G.'tHooft, these proceedings.

\bibitem{gross}  See D.Gross, these proceedings.

\bibitem{lp}Excellent recent analyses are: P. Langacker and N.
Polonsky, Phys. Rev. D 47 (1993) 4028;\\ ibid., D 49 (1994)
1454;\\  L.J. Hall and U. Sarid, Phys. Rev. Lett. 70 (1993) 2673.

\bibitem{swogu}The Second Workshop on Grand Unification, University
of Michigan, Ann Arbor, April 24-26, 1981, eds. J.Leveille, L.Sulak,
D.Unger; Birkhauser, 1981.

\bibitem{ws} S. Weinberg, Phys. Rev. D 13 (1976) 974; D 19 (1979)
1277;\\ L.
Susskind, Phys. Rev. D 20 (1979) 2619.

\bibitem{etc} S. Dimopoulos and L. Susskind, Nucl. Phys. B 155 (1979)
237
;\\ E. Eichten and K. Lane, Phys. Lett. B 90 (1980) 125.

\bibitem{je} S. Dimopoulos, S. Raby and P. Sikivie, Nucl. Phys. B
219, (1982) 479;\\ S. Dimopoulos and J. Ellis, Nucl. Phys. B 182
(1981) 505.

\bibitem{fayet}P. Fayet, Phys. Lett. B 69 (1977) 489;\\ B 84 (1979)
416.

\bibitem{fgp}S. Ferrara, L. Girardello, and F. Palumbo, Phys. Rev.
D 20 (1979) 403.

\bibitem{drw1} S. Dimopoulos, S. Raby, and F. Wilczek, Phys.
Rev. D 24 (1981) 1681.

\bibitem{dg} S. Dimopoulos and H. Georgi, ``Supersymmetric
GUTs'', p 285, Second Workshop on Grand Unification, University of
Michigan, Ann Arbor, April 24-26, 1981, eds. J.Leveille, L.Sulak,
D.Unger; Birkhauser, 1981;\\
S. Dimopoulos and H. Georgi, Nucl. Phys. B 193 (1981) 150.

\bibitem{gg} L. Girardello and M.T. Grisaru, Nucl. Phys. B 194 (1982)
65.

\bibitem{sin2}N. Sakai, Zeit.Phys. C 11 (1981) 153;\\ L. Iba\~nez and
G.G.
Ross, Phys. Lett. B 105 (1981) 439;\\ M. B. Einhorn and D.
R. T. Jones, Nucl. Phys. B 196 (1982) 475;\\ W. J. Marciano
and G. Senjanovic, Phys. Rev. D 25 (1982) 3092.

\bibitem{wsy}S. Weinberg, Phys. Rev. D 26 (1982) 287;\\
N. Sakai and T. Yanagida, Nucl. Phys. B 197 (1982) 533.

\bibitem{drw2} S. Dimopoulos, S. Raby, and F. Wilczek, Phys.
Lett. B 112 (1982) 133;\\ J. Ellis, D.V. Nanopoulos, and S. Rudaz,
Nucl. Phys. B 202 (1982) 43.

\bibitem{dw} S. Dimopoulos and F. Wilczek, Santa Barbara preprint,
July 1981; Proceedings Erice Summer School, Ed. A. Zichichi (1981).

\bibitem{gr}B. Grinstein, Nucl. Phys. B 206 (1982) 387;\\ R.N. Cahn,
I. Hinchliffe, and L. Hall, Phys. Lett. B 109 (1982) 426;\\ A.
Masiero,
D.V. Nanopoulos, K. Tamvakis, and T. Yanagida, Phys. Lett. B 115
(1982)
380;\\
K.S. Babu and S.M. Barr, Phys. Rev D 48 (1993) 5354;\\ D 50 (1994)
3529.

\bibitem{hs}  M. Dine, W. Fischler, Nucl. Phys. B 204 (1982) 346;\\
S. Dimopoulos and S. Raby, Nucl. Phys. B 219 (1982) 479;\\
J. Polchinski and L. Susskind, Phys. Rev. D 26 (1982) 3661.

\bibitem{w}E. Witten, Phys. Lett. B 105 (1981) 267.

\bibitem{suGUT} E. Cremmer, S. Ferrara, L. Girardello, and A. Van
Proyen,
Phys. Lett. B  116 (1982)  231;\\
A. Chamseddine, R. Arnowitt, and P. Nath, Phys. Rev. Lett.
 49 (1982) 970;\\
R. Barbieri, S. Ferrara, and C. Savoy, Phys. Lett. B 110
(1982) 343;\\
L. J. Hall, J. Lykken, and S. Weinberg, Phys. Rev. D 27 (1983)
2359.

\bibitem{rewsb}  L.E. Iba\~nez and G.G. Ross, Phys. Lett. B 110B
(1982) 215;\\ L.~Alvarez-Gaum\'e, M.~Claudson, and M.B.~Wise,
Nucl. Phys. B 207 (1982) 96;\\ M. Dine and W. Fischler, Nucl. Phys. B
204 (1982) 346;\\
K.~Inoue, A.~Kakuto, H.~Komatsu, and S.~Takeshita, Prog. Theor. Phys.
68 (1982) 927 and 71 (1984) 413;\\
J.~Ellis, D.V.~Nanopoulos, and K.~Tamvakis, Phys. Lett. B 121 (1983)
123;\\L. Alvarez-Gaum\'e, J. Polchinski, and M. Wise, Nucl.
Phys. B 221 (1983) 495;\\ L.E. Iba\~nez and C. Lopez, Phys.
Lett. B 126 (1983) 54;\\Nucl. Phys. B 233 (1984) 511;\\
C. Kounnas, A.B. Lahanas, D.V. Nanopoulos, and M. Quiros, Nucl.
Phys. B 236 (1984) 438;\\  L.E. Iba\~nez, C. Lopez, and C. Munoz,
Nucl. Phys. B 256 (1985) 218;\\
G.~Gamberini, G.~Ridolfi, and F.~Zwirner, Nucl. Phys. B 331 (1990)
331.

\bibitem{cpw} M. Carena, S. Pokorski, and C.E.M. Wagner,
Nucl. Phys. B 406 (1993) 59.

\bibitem{fg} P.H. Frampton and S.L. Glashow, Phys. Lett. B 131 (1983)
340, Erratum B 135 (1984) 515;\\ A. Giveon, L.J. Hall, and U. Sarid,
Phys. Lett B 271 (1991) 138.

\bibitem{phystoday} S. Dimopoulos, S. Raby, and F. Wilczek, Phys.
Today 44 (1991)  25-33.

\bibitem{barb}R. Barbieri, G. Dvali, and A. Strumia, Pisa preprint:
IFUP-PTH-94-22;

\bibitem{strGUT}G. Aldazabal, A. Font, L.E. Iba\~nez, and A.M.
Uranga, Madrid preprint:
FTUAM-94-28;\\
S. Chadhouri, S.-W. Chung, and J.D. Lykken, Fermilab-pub-94-137-T;\\
G.Cleaver, OSU preprint.

\bibitem{bg} J. Ellis, K. Enqvist, D.V. Nanopoulos, and F.Zwirner,
Mod. Phys. Lett. A 1 (1986) 57;\\ R. Barbieri and G.F. Giudice,
Nucl. Phys. B 306 (1988) 63.

\bibitem{Higgs}Y. Okada, M. Yamaguchi, and T. Yanagida,
Prog. Theor. Phys. Lett. 85 (1991) 1;\\
J. Ellis, G. Ridolfi, and F. Zwirner, Phys. Lett. B 257 (1991) 83
and B 262 (1991) 477;\\
H.E. Haber and R. Hempfling, Phys. Rev. Lett. 66 (1991)
1815;\\ R.Barbieri and M. Frigeni, Phys. Lett. B 258 (1991) 395.

\bibitem{hkr}L.J. Hall, V. A. Kostelecky, and S. Raby, Nucl.
Phys. B 267 (1986) 415.

\bibitem{bh}R. Barbieri and L.J. Hall, LBL-36022 (1994).

\bibitem{dh}S.Dimopoulos and L.J.Hall,  LBL-36040 (1994).

\bibitem{g} H.Georgi, Phys. Lett. B 169 (1986) 231.






\end{thebibliography}
\end{document}